\documentclass[12pt]{article}

\usepackage[utf8]{inputenc}
\usepackage{cite}
\usepackage{epsfig}
\usepackage{amsmath}
\usepackage{amssymb} %math

\usepackage{array}
\usepackage{multirow}
\usepackage{rotating}
\usepackage{graphicx}
\usepackage{slashed}
\usepackage{float}

\def\mathswitch#1{\relax\ifmmode#1\else$#1$\fi}
\def\mathswitchr#1{\relax\ifmmode{\mathrm{#1}}\else$\mathrm{#1}$\fi}
\newcommand{\PW}{\mathswitchr W}
\newcommand{\PZ}{\mathswitchr Z}

\newcommand{\Pt}{\mathswitchr t}

\newcommand{\MW}{\mathswitch {M_\PW}}
\newcommand{\MZ}{\mathswitch {M_\PZ}}

\newcommand{\scrs}{\scriptscriptstyle}
\newcommand{\sw}{\mathswitch {s_{\scrs\PW}}}
\newcommand{\cw}{\mathswitch {c_{\scrs\PW}}}
\newcommand{\mw}{\mathswitch {\overline{M}_\PW}}
\newcommand{\mz}{\mathswitch {\overline{M}_\PZ}}

\newcommand{\as}{\alpha_{\mathrm s}}
\newcommand{\seff}[1]{\sin^2\theta_{\rm eff}^{#1}}
\newcommand{\msbar}{$\overline{\mbox{MS}}$}

\newcommand{\gev}{\,\, \mathrm{GeV}}
\newcommand{\mev}{\,\, \mathrm{MeV}}
\newcommand{\re}{\text{Re}\,}
\newcommand{\im}{\text{Im}\,}

\newcommand{\mycaption}[1]{\caption{\sl #1}}

\makeatletter
\def\section{\@startsection {section}{1}{\z@}{-3.5ex plus -1ex minus 
 -.2ex}{2.3ex plus .2ex}{\large\bf\boldmath}}
\def\subsection{\@startsection{subsection}{2}{\z@}{-3.25ex plus -1ex
 minus -.2ex}{1.5ex plus .2ex}{\normalsize\bf\boldmath}}
\def\subsubsection{\@startsection{subsubsection}{3}{\z@}{-3.25ex plus
 -1ex minus -.2ex}{1.5ex plus .2ex}{\normalsize\it}}

\graphicspath{{.}{plots/}}

\begin{document}
\thispagestyle{empty}

\def\thefootnote{\fnsymbol{footnote}}

\begin{flushright}
\end{flushright}

\vspace{1cm}

\begin{center}

{\Large {\bf Mixed EW--QCD leading fermionic three-loop corrections at \boldmath $\mathcal{O}(\as\alpha^2)$ to electroweak precision
observables}}
\\[3.5em]
{\large
Lisong Chen and Ayres~Freitas
}

\vspace*{1cm}

{\sl
Pittsburgh Particle-physics Astro-physics \& Cosmology Center
(PITT-PACC),\\ Department of Physics \& Astronomy, University of Pittsburgh,
Pittsburgh, PA 15260, USA
}

\end{center}

\vspace*{2.5cm}

\begin{abstract} Measurements of electroweak precision observables at future electron-position colliders, such as the CEPC, FCC-ee, and ILC, will be sensitive to physics at multi-TeV scales. To achieve this sensitivity, precise predictions for the Standard Model expectations of these observables are needed, including corrections at the three- and four-loop level. In this article, results are presented for the calculation of a subset of three-loop mixed electroweak-QCD corrections, stemming from diagrams with a gluon exchange and two closed fermion loops. The numerical impact of these corrections is illustrated for a number of applications: the prediction of the
W-boson mass from the Fermi constant, the effective weak mixing angle, and the
partial and total widths of the Z boson. Two alternative renormalization schemes for the top-quark mass are considered, on-shell and $\overline{\mbox{MS}}$.
 \end{abstract}

\setcounter{page}{0}
\setcounter{footnote}{0}

\newpage

%%%%%%%%%%%%%%%%%%%%%%%%%%%%%%%%%%%%%%%%%%%%%%%%%%%%%%%%%%%%%%

\section{Introduction}

The term electroweak precision observable (EWPO) refers to a class of quantities that are intimately connected to properties of the electroweak $W$ and $Z$ bosons. These quantities can be measured with high precision, and they permit a clean and robust theoretical description within the Standard Model (SM), with negligible non-perturbative contributions. Therefore they can be used to put severe constraints on many models beyond the SM. A widely used set of EWPOs includes (a) the $W$-boson mass, $M_W$, which can be predicted within the SM from the Fermi constant for muon decay, $G_\mu$, (b) the partial widths of the $Z$-boson into different fermion final states, $\Gamma_f$, extracted from measurements of the cross-section for $e^+ e^- \to f\bar{f}$ at the $Z$ peak, and (c) the effective weak mixing angle $\sin^2{\theta_{eff}^f}$, which is related to the ratio of the vector/axial-vector $Zff$ couplings and can be determined from measurements of parity-violating asymmetries at the $Z$ peak.
Predictions for these observables within the SM are currently known up to full two-loop level\cite{qcd2,mwshort,mwlong,mw,mwtot,swlept,swlept2,swbb,gz,zbos}, and partial three- and four-loop level, where ``partial'' refers to contributions that are enhanced by powers of the top Yukawa coupling, $\alpha_t=\frac{y_t}{4\pi}$ \cite{qcd3,mt6,qcd4}.

It has been proposed that future high-luminosity $e^+e^-$ colliders may re-measure these EWPOs with significantly increased precision. These proposals include the circular colliders CEPC\cite{cepc} and FCC-ee \cite{fccee}, as well as the linear collider ILC/Giga-Z~\cite{Fujii:2019zll}. Hence the further effort from the theory side is desirable. In Ref.~\cite{therr}, it has argued that the most important missing higher-order corrections are the three-loop contributions given at $\mathcal{O}(\alpha^3)$, $\mathcal{O}(\alpha^2\as)$, $\mathcal{O}(\alpha\as^2)$, where $\alpha$ and $\as$ refer to EW and QCD couplings, respectively. These corrections can be further classified according to the number of closed fermion loops in the diagrams. Contributions with multiple fermion loops are especially important since they are numerically enhanced by powers of the top-quark mass and the total flavor number of fermion flavors.  

In \cite{Chen:2020xzx}, results have been presented for the three-loop electroweak corrections with the maximal number of fermion loops.
In this article, we report on the computation of the leading fermionic correction at the order $\mathcal{O}(\alpha^2\as)$, which involves diagrams with two closed fermion loops and one gluon exchange. It is demonstrated that the size of the leading fermionic corrections at $\mathcal{O}(\alpha^2\as)$ are comparable to those at the order $\mathcal{O}(\alpha^3)$, as expected from Ref.~\cite{therr}. 

This paper is organized as follows. In section~\ref{renorm}, we introduce the renormalization schemes we adopted and illustrate the types of diagrams relevant for the mixed EW--QCD contributions considered in this calculation. Section~\ref{obs} discusses the definition of each EWPOs that we compute. We describe the methods used for the calculation, with a discussion of several technique aspects, in section~\ref{calc}. Numerical results for each EWPO are given in section~\ref{res1} and section~\ref{res2} for two different definitions on the top-quark mass: on-shell and \msbar. Finally, our summary is given in section~\ref{conc}.

%%%%%%%%%%%%%%%%%%%%%%%%%%%%%%%%%%%%%%%%%%%%%%%%%%%%%%%%%%%%%%

\section{Renormalization}
\label{renorm}

In this work, the on-shell (OS) renormalization scheme is adopted for electroweak radiative corrections. However, when computing $\mathcal{O}(\alpha^2\as)$ corrections to electroweak observables, one needs to include QCD corrections in the renormalization of the top-quark mass, and the modified minimal subtraction (\msbar) scheme is more commonly used for QCD calculations. Therefore, we consider two alternative schemes for the top-quark mass, OS, and \msbar.
The OS mass is closely related to the experimental top-quark mass determined by using the template fit approach \cite{Buckley:2011ms}. However, the OS top mass is subject to the renormalon ambiguity and other non-perturbative QCD effects. In contrast, the \msbar\ mass is protected from such long-distance phenomena and thus preferable from a theoretical point of view. Therefore, it is worth to perform the calculation in both mass renormalization schemes. The values of the top-quark mass in these two schemes are related by a non-divergent function, which has been computed up to four-loop level \cite{osmsbar}.

\medskip

In the on-shell scheme, the electromagnetic charge is defined as the electromagnetic coupling strength in the Thomson limit, and on-shell external fields are renormalized to unity. Furthermore, renormalized masses are defined through the location of the propagator poles. However, for an unstable massive particle, the pole of the propagator is complex, in which case the renormalized mass coincides with the real part of the complex pole, while the imaginary part is associated with the decay width of the particle as follows,
\begin{align}
s_0 \equiv \overline{M}^2 - i\overline{M}\overline{\Gamma}, \label{cplxm}
\end{align}
where $\overline{M}$ is the on-shell mass, while $\overline{\Gamma}$ is the
decay width of an unstable massive particle\footnote{\label{massdef}We would like to remind the reader that this definition of the mass and width is theoretically well-defined and gauge-invariant \cite{zpole}, but it differs from the definition used in experimental analyses. Due to this difference, the experimental mass and width $M$, $\Gamma$, respectively, are related to $\overline{M}$, $\overline{\Gamma}$ by the relations 
$\overline{M} = M\big/\sqrt{1+\Gamma^2/M^2},$
$\overline{\Gamma} = \Gamma\big/\sqrt{1+\Gamma^2/M^2}$ \cite{mrel}.}.

Including radiative corrections, the inverse massive gauge boson two-point function can be written as
\begin{equation}
D(p^2)  = p^2 - \overline{M}^2 -\delta Z(p^2-\overline{M}^2)+ \Sigma(p^2) - \delta \overline{M}^2, \label{bosprop}
\end{equation}
where $\Sigma(p^2)$ is the transverse part of the gauge boson self-energy,
and $\delta \overline{M}^2$ is the mass counterterm. Quantity $\delta Z$ describes the counterterm contributions of the field strength. For simplicity we set $ \delta Z=0$ for the massive gauge bosons, which is justified by the fact that, being unstable, they appear only as internal particles in a physical
process\footnote{In our calculations, we have checked
explicitly that any field renormalization counterterms for the massive gauge bosons cancel in the result for a physical observable.}.

\begin{figure}[tb]
\centering
\begin{equation}
\begin{split}
\Sigma_{t(\as)} & = \quad
\raisebox{-8pt} {\includegraphics[width=0.13\linewidth]{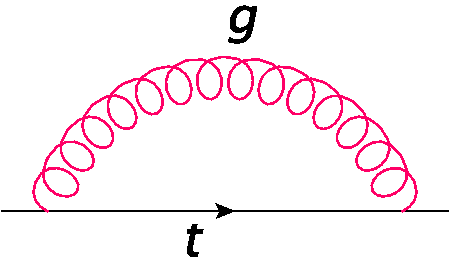}}\\
\Sigma_{V_1V_2(\as \alpha)} & =
\raisebox{-33pt} {\includegraphics[width=0.8\linewidth]{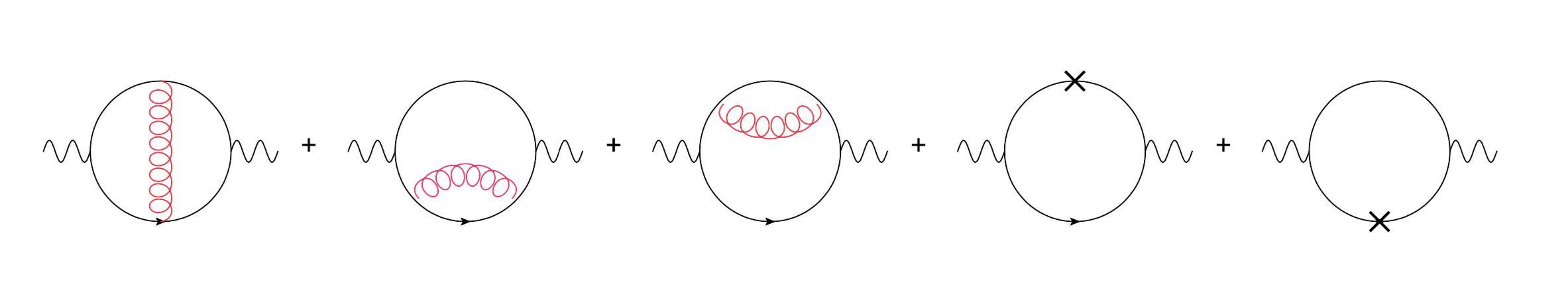}}\\
\Sigma_{V_1V_2(\as\alpha^2)} & = \quad
\raisebox{-17pt} {\includegraphics[width=0.63\linewidth]{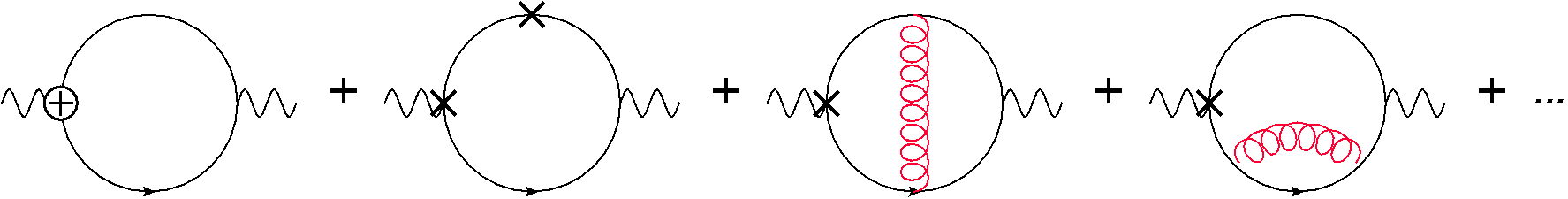}}
\end{split}  \notag
\end{equation}
\vspace{-2em}
\mycaption{Diagrams with closed fermion loops contributing to self-energies at
different orders. $V_1$ and $V_2$ denote the possible different in- and outgoing gauge bosons. Vertices "$\oplus$" and "$\times$" indicate the counterterms at the loop order $\mathcal{O}(\as\alpha)$  and $\mathcal{O}(\alpha)$ or $\mathcal{O}(\as)$, respectively. Note that there are no actual three-loop diagrams with two explicit closed fermion loops at the order $\mathcal{O}(\as\alpha^2)$, but instead, all contributions stem from sub-loop counterterm insertions.
\label{fig:sesub}}
\end{figure}

In the on-shell scheme, setting $D(s_0)=0$ and using eqs.~\eqref{cplxm} and \eqref{bosprop} leads to the conditions
\begin{align}
\delta\overline{M}^2 &= \re \Sigma\bigl(\overline{M}^2 -
i\overline{M}\overline{\Gamma}\bigr), \label{dmdef} \\
\overline{\Gamma} &= \frac{1}{\overline{M}} \, \im \Sigma\bigl(\overline{M}^2 -
i\overline{M}\overline{\Gamma}\bigr). \label{gamdef} 
\end{align}
Expanding eq.~\eqref{dmdef} and eq.~\eqref{gamdef} iteratively in
orders of perturbation theory, the $W$-mass counterterm is found to be 
\begin{equation}
\begin{split}
\delta \overline{M}^2_{\PW(\as \alpha)}  & = \re \Sigma_{\PW(\as \alpha)}(\mw^2), \\ 
\delta \overline{M}^2_{\PW(\as \alpha^2 )} & = \re \Sigma_{\PW(\as \alpha^2)}(\mw^2) + [\im  \Sigma_{\PW(\as \alpha)}(\mw^2)]\, [\im  \Sigma'_{\PW(\alpha)}(\mw^2)]+ [\im  \Sigma'_{\PW(\as \alpha)}(\mw^2)]\, [\im  \Sigma_{\PW(\alpha)}(\mw^2)].
\end{split} \label{countermw}
\end{equation}
%\end{align}
Here and in the following the subscripts in parenthesis denote the loop order. Furthermore, $\Sigma'$ denotes the derivative of the self-energy with respect to the external momentum.
Due to the $\gamma$--$Z$ mixing effects, deriving the $Z$-mass counterterm is more subtle (see detailed discussion in Ref.~\cite{Chen:2020xzx}):
\begin{equation}
\begin{split}
\delta \overline{M}^2_{\PZ(\as \alpha)} & = \re \Sigma_{\PZ\PZ(\as \alpha)}(\mz^2)\,,\\
\delta \overline{M}^2_{\PZ(\as \alpha^2)} &= \re \Sigma_{\PZ\PZ(\as \alpha^2)}(\mz^2) +
[\im  \Sigma_{\PZ\PZ(\as \alpha)}(\mz^2)] \, [\im  \Sigma'_{\PZ\PZ(\alpha)}(\mz^2)] \\  
&\quad +[\im  \Sigma_{\PZ\PZ(\alpha)}(\mz^2)] \, [\im  \Sigma'_{\PZ\PZ(\as \alpha)}(\mz^2)] \\                                                              &\quad +\frac{2}{\mz^2} [\im  \Sigma_{\gamma\PZ(\as \alpha)}(\mz^2)] \,  [\im  \Sigma_{\gamma\PZ(\alpha)}(\mz^2)] + \frac{1}{2}\mz^2 \,\delta Z^{\gamma\PZ}_{(\alpha)}\,\delta Z^{\gamma\PZ}_{(\as\alpha)}\,.
\end{split} \label{countermz}
\end{equation}
The self-energies receive contributions from one and two-loop diagrams with counterterm
insertions shown in Fig.~\ref{fig:sesub}.

For counterterm insertions on internal fermionic lines, one needs to derive the mass counterterms for massive fermions. Writing the inverse massive fermion two-point function as
\begin{equation}
\begin{split}\label{}
D_{\psi}(p)&= \slashed{p}-M+\Sigma_{\psi}(p^2)+\delta Z_{\psi} (\slashed{p}-M)-\delta M_{\psi} \\
&=\slashed{p}[1+\Sigma_V(p^2)+\delta Z_{\psi}]+M\Bigl[-1+\Sigma_S(p^2)-\delta Z_{\psi}-\frac{\delta M}{M}\Bigr],
\end{split}
\end{equation}
where $\Sigma_{\psi}(p^2) = \slashed{p}\Sigma_V(p^2) + M\Sigma_S(p^2)$ is the fermion self-energy, which can be split into a vector and a scalar part\footnote{Here we have ignored parity-violating interactions since for our purposes we only need QCD corrections to the fermion masses.}.
In this work, the masses of all fermions except the top quark is set to zero. 
The OS top mass is defined by 
imposing the condition $D_t(p)|_{p^2=M_t^2-iM_t\Gamma_t}=0$ and expanding up to one-loop order (in which case the top width can be neglected), leading to
\begin{equation}
\delta M_{t(\as)}=M_t [\re \Sigma_{V(\as)}(M_t^2)+\re \Sigma_{S(\as)}(M_t^2)].
\end{equation}
Alternatively, in the \msbar\ scheme the counterterm only contains  the divergent piece along with $\log{4\pi}$ and the Euler number $\gamma_E$. At one-loop order it reads\footnote{In our practical calculations, we excluded the dependence on $\log{4\pi}$ and $\gamma_E$ to match the conventions used in the program {\sc TVID~2.1}\cite{Bauberger:2019heh}, which we use for the evaluation of master integrals (see section~\ref{calc} for more details). }
\begin{equation}
\delta m_{t(\as)}=-\frac{3 C_F g_s^2}{16 \pi^2}\Bigl(\frac{1}{\epsilon}+\log{4\pi}-\gamma_E\Bigr)\,m_t(\mu).
\label{msbtop}
\end{equation}
Here the lower case $m$ is used to denote \msbar\ quantities, and $\mu$ is the renormalization scale.

At one-loop level, the relation between the OS and \msbar\ mass can be easily derived from these formulae, with the result
\begin{equation}
\frac{M_t}{m_t}=1+\frac{\as C_F}{4\pi} \Bigl(3\log{\frac{M^2_t}{\mu^2}-4}\Bigr)+\mathcal{O}(\as^2). \label{massrel}
\end{equation}

\medskip

%-------------------------------------------------------------
%-------------------------------------------------------------

The weak mixing angle counterterm is obtained from the $W$ and $Z$ mass counterterms, see Ref.~\cite{Chen:2020xzx}, while the $\delta Z^{\PZ\gamma}_{(n)}$ turns out to be zero at all order needed in this work.
Finally, the charge renormalization is given by
\begin{align}
\delta Z_{e(\alpha)} &= \frac{1}{2}\bigl[\Delta\alpha+\Sigma_{\gamma\gamma(\alpha)}^{\rm top\,\prime}(0)+\Pi_{\gamma\gamma(\alpha)}^{\rm lf}(\MZ^2)\bigr], \label{dze1} \\
\delta Z_{e(\as\alpha)} &= \frac{1}{2}\bigl[\Sigma_{\gamma\gamma(\as\alpha)}^{\rm top\,\prime}(0)+\Pi_{\gamma\gamma(\as\alpha)}^{\rm lf}(\MZ^2)\bigr], \label{dze2} \\
\delta Z_{e(\as\alpha^2)} &= 3\,\delta Z_{e(\alpha)}\delta Z_{e(\as\alpha)}, \label{dze3} 
\intertext{where}
\Pi_{\gamma\gamma}(q^2) &=  \frac{\Sigma_{\gamma\gamma}(q^2)}{q^2}\,. 
\end{align}
Here $\Sigma_{\gamma\gamma}^{\rm top\,\prime}$ denotes the derivative of the massive-top loop contribution to the photon self-energy. The symbol $\Delta\alpha$ in eq.~\eqref{dze1} stems from light-fermion two-loop contributions in the photon vacuum polarization,
\begin{equation}
\Delta\alpha =
\Pi_{\gamma\gamma}^{\rm lf}(\MZ^2)-\Pi_{\gamma\gamma}^{\rm lf}(0), \qquad 
\label{delalph}
\end{equation}
where $\Pi_{\gamma\gamma}^{\rm lf}(q^2)$ can be further divided into a leptonic part, which is
perturbatively calculable \cite{dalept}, and a hadronic part that becomes non-perturbative for
small $q^2$. With the help of dispersion relation, one can extract the hadronic contribution from experimental measurements of $e^+e^- \to \text{had.}$ \cite{dahad}. Given that $\Delta\alpha$ is inherently non-perturbative, it is not strictly associated with any loop order in \eqref{dze1}--\eqref{dze3}, but one can maintain the correct book-keeping by including it in the one-loop counterterm. 

%%%%%%%%%%%%%%%%%%%%%%%%%%%%%%%%%%%%%%%%%%%%%%%%%%%%%%%%%%%%%%

\section{Computation of observables}
\label{obs}

In the SM the Fermi constant is defined through
\begin{equation}
G_\mu = \frac{\pi\alpha}{\sqrt{2}\sw^2\mw^2}(1+\Delta r), \label{gmu}
\end{equation}
where $\sw^2=1-\mw^2/\mz^2$ and $\Delta r$ includes the contribution from radiative corrections. When focusing on corrections with maximal number of closed fermion loops, $\Delta r$ receives contributions from the $W$-boson self-energy and counterterms for the $W$ mass and $W\ell\nu_\ell$ vertex, see Ref.~\cite{Chen:2020xzx} for details.

$G_\mu$ can be determined with high precision from measurements of the muon decay lifetime\cite{Tishchenko:2012ie} after subtracting QED corrections~\cite{vanRitbergen:1999fi}. This measurement together with eq.~\eqref{gmu} can then be used to obtain a prediction for the $W$-boson mass within the SM.

\bigskip\noindent

The effective weak mixing angle and the partial decay widths are both related to the effective vector and axial-vector couplings of the $Z$-boson to $f\bar{f}$, denoted $v_f$ and $a_f$, respectively. When limiting ourselves to corrections with closed fermion loops, $v_f$ and $a_f$ receive contribution from vertex counterterms, as well as photon--Z mixing. See Ref.~\cite{Chen:2020xzx} for explicit expressions.

In terms of these effective couplings, the effective weak mixing angle is given by
\begin{equation}
\seff{f} = \frac{1}{4|Q_f|}\Bigl(1+\re\frac{v_f}{a_f}\Bigr)_{s=\mz^2}\,.
\end{equation}
The derivation of the partial width, $\overline{\Gamma}_f$, for $Z \to f\bar{f}$ is more involved and requires the use of the optical theorem. Its has the general form
\begin{align}
\overline{\Gamma}_f = \frac{N_c^f\mz}{12\pi} C_\PZ \Bigl [
 {\cal R}_{\rm V}^f |v_f|^2 + {\cal R}_{\rm A}^f |a_f|^2 \Bigr ]_{s=\mz^2} 
 \;. \label{Gz}
\end{align}
Here $N_c^f = 3(1)$ for quarks (leptons), and the radiator functions ${\cal R}_{\rm V,A}$ represent final-state QCD and QED corrections. For contributions with maximal number of closed fermion loops they are simply 
${\cal R}_{\rm V,A}=1$. $C_Z$ depends on $Z$ self-energy contributions and must be determined recursively order-by-order \cite{Chen:2020xzx}. At ${\cal O}(\alpha^2\as)$ one finds
\begin{align}
\Delta\overline{\Gamma}_{f,(\alpha^2\as)} &= \frac{N_c^f\mz}{12\pi}  \Bigl [
 \Delta F^f_{\rm V,(\alpha^2\as)} + \Delta F^f_{\rm A,(\alpha^2\as)}\Bigr ]_{s=\mz^2} \,,\\
\delta F_{\rm V(\alpha^2\as)}^f &= v_{f(0)}^2 \bigl[  2(\re\Sigma'_{\PZ(\alpha)})(\re\Sigma'_{\PZ(\as \alpha)})-\re\Sigma'_{\PZ(\as \alpha^2)}\notag\\
&\hspace{7.5ex} -\tfrac{1}{2}(\im \Sigma_{\PZ(\alpha)})(\im \Sigma''_{\PZ(\as\alpha)})-\tfrac{1}{2} \delta Z_{\gamma Z}^{(\alpha)}\delta Z_{\gamma _Z}^{(\as\alpha)} \bigr]\notag\\
&\quad +2\,\re (v_{f(0)}v_{f(\alpha)})  (-\re\Sigma'_{\PZ(\as \alpha)})+2\,\re (v_{f(0)}v_{f(\as\alpha)}) (-\re\Sigma'_{\PZ( \alpha)})\notag\\
&\quad +2\,\re (v_{f(\alpha)}^*v_{f(\as\alpha)})+2\, \re(v_{f(0)}v_{f(\as\alpha^2)})
 , \label{Fv} 
\end{align}
where $\delta F_V^f$ contains the contributions from $v_f$ and $C_\PZ$ expanded to the given order, and $\delta F_{\rm A}^f(\as\alpha^2)$ is given analogously by replacing $v_f$ with $a_f$.
Note that $\Sigma_\PZ$ includes $Z$--$\gamma$ mixing effects, given by
\begin{align}
\Sigma_\PZ(p^2) &= \Sigma_{\PZ\PZ}(p^2) 
 - \frac{[\hat{\Sigma}_{\gamma\PZ}(p^2)]^2}{p^2+\hat{\Sigma}_{\gamma\gamma}(p^2)
}
 \label{sigmaz}
 ,\\
\hat{\Sigma}_{\gamma\PZ}(p^2) &= \Sigma_{\gamma\PZ}(p^2) +
 \tfrac{1}{2}\delta Z^{\PZ\gamma}(p^2-\mz^2 - \delta\mz^2)
 + \tfrac{1}{2}\delta Z^{\gamma\PZ}p^2, \\
\hat{\Sigma}_{\gamma\gamma}(p^2) &= \Sigma_{\gamma\gamma}(p^2)
 + \tfrac{1}{4}(\delta Z^{\PZ\gamma})^2(p^2-\mz^2 - \delta\mz^2).
\end{align}
Furthermore, when neglecting light fermion masses, $\im \Sigma''_\PZ = 0$ at 1- and 2-loop order.

\section{Technical aspects of the calculation} 
\label{calc}

%-------------------------------------------------------------
\begin{figure}[ht!]
\centering
\includegraphics[width=0.8\linewidth]{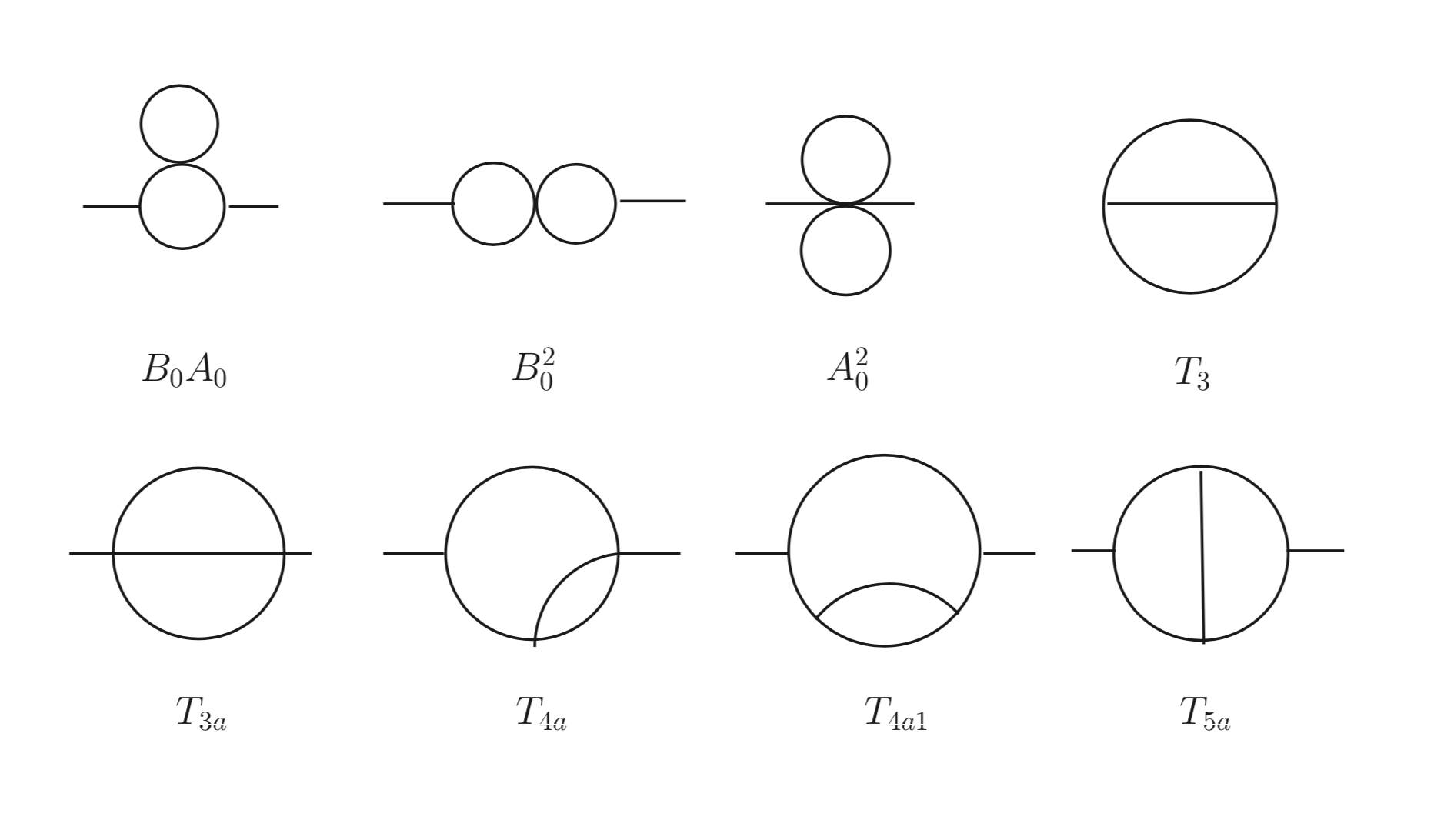}
\vspace{-1.5em}
\mycaption{The master integral basis chosen in our calculation, with notation taken from \cite{Bauberger:2019heh}.\label{fig:masint}}
\end{figure}
%-------------------------------------------------------------
Most algebraic computations in this project are carried out by computer tools due to the lengthy expressions,
within the framework of {\sc Mathematica}. {\sc FeynArts 3.3} \cite{feynarts} and {\sc FeynCalc
9.2.0} \cite{feyncalc} have been employed for generating amplitudes and computing some of the Dirac and tensor algebra. The reduction of two-loop amplitudes to a set of master integrals has been performed with two independent methods, one using integration-by-parts (IBP) identities \cite{Chetyrkin:1981qh} as implemented in 
{\sc FIRE6}\cite{Smirnov:2019qkx}, and the other using the integral reduction techniques of Ref.~\cite{Weiglein:1993hd}. The two-loop master integral topologies used in this work are shown in Fig.~\ref{fig:masint}.
 {\sc TVID 2.1} \cite{Bauberger:2019heh} is used for the numerical evaluation of these master integrals. The masses and Yukawa couplings of all fermions except the top-quark have been
neglected. Furthermore, CKM mixing of the top quark with other quark generations
has been ignored.

\bigskip\noindent

For the computation of the renormalization counterterms, one needs the derivatives of two-loop self-energy functions, see $e.\,g.$
eqs.~\eqref{countermw},\eqref{countermz},\eqref{dze1},\eqref{dze2}. 
The procedure for taking the derivative at zero external momentum is different from the situation with non-zero momentum. Let us define $I(\nu_1,\nu_2,..., m_1,m_2,..,; p^2)$ as one of the two-loop scalar integrals according to
\begin{equation}
\begin{aligned}
&I(\nu_1,\nu_2,..., m_1,m_2,..,; p^2) \\
&\equiv \int \frac{d^D q_1\,d^D q_2}{(q_1^2-m_1^2)^{\nu_1}((q_1+p)^2-m_2^2)^{\nu_2}((q_2-q_1)^2-m_3^2)^{\nu_3}(q_2^2-m_4^2)((q_2+p)^2-m_5^2)^{\nu_5}}\,.
\end{aligned}
\end{equation}
For $p^2=0$, one has \cite{afras}
\begin{equation}
\begin{split}
&\frac\partial{\partial p^2}I(...;p^2=0)=\frac{1}{2D}\frac{\partial^2}{\partial p_{\mu}\partial p^{\mu}}I(...;p^2)\bigg|_{p^2=0}\\
                     &\quad= \frac{2}{D}\biggl[\Bigl(1+\nu_2+\nu_5-\frac{D}{2}\Bigr)(\nu_2 I(\nu_2+1)+\nu_5 I(\nu_5+1))\\
                     &\qquad\quad+m_2^2 \nu_2(\nu_2+1)I(\nu_2+2)+m_5^2\nu_5(\nu_5+1)I(\nu_5+2)\\
                     &\qquad\quad +\nu_2\nu_5((m_2^2-m_3^2+m_5^2)I(\nu_2+1,\nu_5+1)-I(\nu_2+1,\nu_3-1,\nu_5+1))
		     \biggr]_{p^2=0},
                     \end{split}                     
\end{equation}
whereas for $p^2 \neq 0$, one obtains \cite{gz}
\begin{equation}
\begin{split}
\frac{\partial}{\partial p^2} I(...;p^2\neq 0)  &=-\frac{1}{2p^2} p^{\mu} \frac{\partial}{\partial p^{\mu}} I(...; p^2)\\
                    & = -\frac{1}{2 p^2} \bigl[(\nu_2 +\nu_5)I -\nu_2 I(\nu_1-1,\nu_2+1)-\nu_5 I(\nu_4-1,\nu_5 +1)\\
                    &\qquad\quad +\nu_2(m_2^2-m_1^2+p^2)I(\nu_2+1)+\nu_5(m_5^2-m_4^2+p^2)I(\nu_5+1)\bigr].
\end{split}
\end{equation}
Then by using IBP identities, one can reduce the new $I(...;p^2)$ to a linear combination of the chosen master integrals in Fig.~\ref{fig:masint}. This can be carried out, for example, with the help of {\sc FIRE6}.

All steps of the calculations described above have been carried out in two independent implementations. Due to the ambiguity in the choice of master integrals, the final algebraic expressions in terms of these basis functions may not be unique, so that a comparison at this level is difficult. However, numerical comparisons with the help of TVID yielded very good agreement between the two results. The cancellation of UV divergencies in TVID can be checked algebraically.

%%%%%%%%%%%%%%%%%%%%%%%%%%%%%%%%%%%%%%%%%%%%%%%%%%%%%%%%%%%%%%

\section{Numerical results in the on-shell scheme}
\label{res1}

To illustrate the numerical impact of the leading fermionic ${\cal O}(\alpha^2\as)$ corrections, this section presents results for the input parameters listed in Tab.~\ref{tab:input}. The results do not depend very strongly on the choice of input values within experimentally
allowed ranges. The numerical evaluation of master integrals has been carried out with {\sc TVID~2.1}. Some $\mathcal{O}(D-4)$ coefficients from scalar one-loop integrals have been computed by following Eq.~4.1 in Ref.\cite{Nierste:1992wg}.
%-------------------------------------------------------------
\begin{table}[tb]
\renewcommand{\arraystretch}{1.2}
\begin{center}
\begin{tabular}{|r@{$\;=\;$}ll|}
\hline
$\MZ$ & $91.1876\gev$ & \multirow{2}{*}{$\biggr\}\!\Rightarrow\;
 \mz = 91.1535\gev$} \\
$\Gamma_\PZ$ & $2.4952\gev$ & \\
$\MW$ & $80.358\gev$ & \multirow{2}{*}{$\biggr\}\!\Rightarrow\;
 \mw = 80.331\gev$} \\
$\Gamma_\PW$ & $2.089\gev$ & \\
$M_t$ & $173.0\gev$ & \\
$M_{f\neq \Pt}$ & 0 & \\
$\as$ & 0.1179 & \\
$\alpha$ & \multicolumn{2}{@{}l|}{$1/137.035999084$} \\
$\Delta\alpha$ & $0.05900$ & \\
$G_\mu$ & \multicolumn{2}{@{}l|}{$1.1663787 \times 10^{-5}\gev^{-2}$} \\
\hline
\end{tabular}
\end{center}
\vspace{-2ex}
\mycaption{Benchmark values for the input parameters used in the numerical
analysis, based on Ref.~\cite{pdg}.
\label{tab:input}}
\end{table}
%%-------------------------------------------------------------

The correction to $\Delta r$ is found to be
\begin{align}
\Delta r_{(\alpha^2\as)} = -0.000109. \label{delros}
\end{align}
Using eq.~\eqref{gmu}, this can be translated into a shift for the predicted value of the $W$-boson mass, given by 
\begin{align}
\Delta \overline{M}_{\PW(\alpha^2\as)} \approx \frac{\pi \alpha \mz^2}{2\sqrt{2}G_\mu \mw
 (\mz^2-2\mw^2)} \,\Delta r_{(\alpha^2\as)}
 = 1.70\mev. \label{dmw}
 \end{align}
For the effective weak mixing angle and partial decay widths one obtains 
\begin{align}
&\Delta\sin^2\theta^f_{\rm eff,(\alpha^2\as)} = 1.31\times 10^{-5}, 
\qquad\qquad [\text{independent of the fermion type } f]\\[1ex]
&\begin{aligned}[c]
&\Delta\overline{\Gamma}_{f,(\alpha^2\as)} = N_c^f \bigl[ -7.9\times 10^{-4}\, (I_3^f)^2
 + 2.69\times 10^{-3} \, I_3^f Q_f - 0.0168 \, Q_f^2 \bigr] \mev, \\
&\Delta\overline{\Gamma}_{\ell,(\alpha^2\as)} = -0.0157\mev, \\
&\Delta\overline{\Gamma}_{\nu,(\alpha^2\as)} = -2.0\times 10^{-4}\mev, \\
&\Delta\overline{\Gamma}_{\rm d,(\alpha^2\as)} = -0.0049\mev, \\
&\Delta\overline{\Gamma}_{\rm u,(\alpha^2\as)} = -0.0203\mev, \\
&\Delta\overline{\Gamma}_{\rm tot,(\alpha^2\as)} = -0.103\mev.
\end{aligned} \label{dgamfx} 
\end{align}
In these results, $\MW$ is taken as an independent input parameter. However, one can also assume that $\MW$ is predicted from $G_\mu$, in which case one needs to include the leading effect of
the shift $\Delta M_{\PW(\alpha^2\as)}$ from eq.~\eqref{dmw}, yielding
\begin{align}
&\Delta'\sin^2\theta^f_{\rm eff,(\alpha^2\as)} = 
 \Delta\sin^2\theta^f_{\rm eff,(\alpha^2\as)} - \frac{2\Delta \overline{M}_{\PW(\alpha^2\as)}\mw}{\mz^2}
 = -1.98 \times 10^{-5}, \label{dseff} \\[1ex]
&\Delta'\overline{\Gamma}_{f,(\alpha^2\as)} = 
 \Delta\overline{\Gamma}_{f,(\alpha^2\as)} - \frac{2\Delta\overline{M}_{\PW(\alpha^2\as)}\mw}{\mz}
 \times \frac{\alpha N_c^f}{6\sw^4\cw^4} \bigl[(2\sw^2-1)(I_3^f)^2
  + 2\sw^4Q_f(Q_f-I_3^f)\bigr] \notag \\
&\phantom{\Delta'\overline{\Gamma}_{f,(\alpha^2\as)}} = N_c^f \bigl[0.0663\, (I_3^f)^2
 +0.0148 \, I_3^f Q_f - 0.0289 \, Q_f^2 \bigr] \mev, \\
&\begin{aligned}[c]
&\Delta'\overline{\Gamma}_{\ell,(\alpha^2\as)} = -0.0049\mev, \\
&\Delta'\overline{\Gamma}_{\nu,(\alpha^2\as)} = 0.0166\mev, \\
&\Delta'\overline{\Gamma}_{\rm d,(\alpha^2\as)} = 0.0475\mev, \\
&\Delta'\overline{\Gamma}_{\rm u,(\alpha^2\as)} = 0.0260\mev, \\
&\Delta'\overline{\Gamma}_{\rm tot,(\alpha^2\as)} = 0.2296\mev.
\end{aligned} \label{dgamf} 
\end{align}
The results presented above have been computed in terms of the gauge-invariant complex-pole definitions of the gauge-boson masses and widths, see the footnote on page~\pageref{massdef}. However, when translating the conventional (unbarred) mass and width definition, the results to not change within the significant digits quoted above, since $\Gamma^2/M^2$ is less than $10^{-3}$ for the $W$ and $Z$ bosons.

Compared with the current experimental precision for the EWPOs listed above, mainly from LEP, SLD and LHC, the mixed EW--QCD fermionic three-loop corrections computed here are negligible. For
example, the direct measurements of the $W$ mass, effective weak mixing angle,
and $Z$ width are $\delta\MW^{\rm exp}=0.012\gev$, $\delta\seff{\ell,\rm exp}= 0.00016$ and $\delta\Gamma_{\PZ,\rm
tot}^{\rm exp}= 0.0023\gev$. These are at least one order of magnitude larger
than the corrections in eqs.~\eqref{dmw}, \eqref{dseff} and \eqref{dgamf}. 
However, for the anticipated precision of future high-luminosity $e^+e^-$ colliders, such as FCC-ee, CEPC or ILC \cite{cepc,fccee,Fujii:2019zll}, the corrections computed in this paper cannot be ignored, see Tab.~\ref{tab:futcoll}.

In particular, when combining with the recently computed leading fermionic electroweak three-loop corrections \cite{Chen:2020xzx}, see Tab.~\ref{tab:numcomp}, we observe significant corrections for $\Delta\mw$ and $\Delta'\overline{\Gamma}_\PZ$, while $\Delta' \seff{}$ is rather small due to an accidental cancellation.

%-------------------------------------------------------------
\begin{table}[tb]
\renewcommand{\arraystretch}{1.2}
\centering

\begin{tabular}{|l | c | c | c |}\hline
                              & CEPC & FCC-ee         & ILC/GigaZ \\
                         \hline
$\MW$[MeV]  &$1$        & $1$                     & $2.5$   \\
$\Gamma_Z$[MeV]  & $0.5$       & $0.1$ &           $1.0$    \\
$\seff{f}$ [$10^{-5}$] &  $2.3$      &          $0.6$               &             $1$  \\
\hline
\end{tabular}
\mycaption{This table demonstrates the future experimental
accuracies projected for CEPC, FCC-ee, and ILC for three EWPOs~\cite{cepc,fccee,Fujii:2019zll,Wilson:2016hne}. For ILC, the GigaZ option is considered, which is a $Z$-pole run with 100~fb$^{-1}$.}
\label{tab:futcoll}
\end{table}
%-------------------------------------------------------------
\begin{table}[tb]
\renewcommand{\arraystretch}{1.2}
\begin{tabular}{|l | r | r | r | r | r |}\hline
                              & $\Delta \mw$ (MeV) & $\Delta \seff{}\;$ & $\Delta' \seff{}\;$ & {$\Delta\overline{\Gamma}_{\rm tot}$ [MeV]} & {$\Delta' \overline{\Gamma}_{\rm tot}$ [MeV]} \\ \hline
$\mathcal{O}(\alpha^3)$         & $-$0.389                     & $1.34\times 10^{-5}$   & $2.09 \times 10^{-5}$   & 0.331                                                  & 0.255                                                    \\ 
$\mathcal{O}(\alpha^2\as)$ & 1.703                       & $1.31\times 10^{-5}$   & $-1.98\times 10^{-5}$   & $-$0.103                                                 & 0.229                                                    \\ 
Sum                       & 1.314                      & $2.65\times 10^{-5}$   & $0.11\times 10^{-5}$    & 0.228                                                  & 0.484                                                     \\ \hline
\end{tabular}
\mycaption{This table shows the numerical results of the leading fermionic three-loop corrections to EWPOs at $\mathcal{O}(\alpha^3)$ from Ref.~\cite{Chen:2020xzx} and at $\mathcal{O}(\alpha^2\as)$ presented here. One can see that the two contributions have comparable size, except for $\Delta \overline{M}_W$, where the mixed EW--QCD three-loop correction is about four times larger in magnitude than the pure EW three-loop. }
\label{tab:numcomp}
\end{table}
%-------------------------------------------------------------

%%%%%%%%%%%%%%%%%%%%%%%%%%%%%%%%%%%%%%%%%%%%%%%%%%%%%%%%%%%%%%

\section{Numerical results in terms of the \msbar\ top mass}
\label{res2}

The numerical evaluation of the leading fermionic ${\cal O}(\alpha^2\as)$ corrections in terms of the \msbar\ top mass keeps all input parameters the same given in Tab.~\ref{tab:input}, except the top-quark mass, which we set to 
\begin{equation}
m_t(\mu=m_t)=163.229 \gev. \label{mtmsbar}
\end{equation}
The numerical results are summarized in Tab.~\ref{tab:msbarewpos}.
While they have the same order of magnitude as the OS results in section~\ref{res1}, the specific numerical values differ noticeably. Overall, the fermionic ${\cal O}(\alpha^2\as)$ corrections are smaller in magnitude when using the \msbar\ top-quark  mass rather than the OS mass. This matches the pattern in previous calculations of ${\cal O}(\alpha\as^n)$, where a better convergence behavior was observed for the \msbar\ top mass \cite{mt6,qcd4}.

%---------------------------------------------------------------
\begin{table}[b]
\renewcommand{\arraystretch}{1.2}
\centering
\begin{tabular}[t]{|c|c|}
\hline
$\Delta r_{(\alpha^2\as)}$ [$10^{-4}$] & $\Delta M_{\PW(\alpha^2\as)}$ [MeV] \\
\hline
$-0.50$ & 0.78 \\
\hline
\end{tabular}
\hspace{2em}
\begin{tabular}[t]{|l|l|l|}
\hline
$X$ & $\Delta X_{(\alpha^2\as)}$ & $\Delta' X_{(\alpha^2\as)}$ \\
\hline
$\seff{}$ [$10^{-5}$] & \phantom{$-$}0.75 & $-$0.76 \\
$\Gamma_\ell$ [MeV] & $-$0.0003 & \phantom{$-$}0.0047 \\
$\Gamma_\nu$ [MeV] & \phantom{$-$}0.0009 & \phantom{$-$}0.0086 \\
$\Gamma_{\rm d}$ [MeV] & $-$0.0018 & \phantom{$-$}0.0223 \\
$\Gamma_{\rm u}$ [MeV] & $-$0.0029 & \phantom{$-$}0.0183 \\
$\Gamma_{\rm tot}$ [MeV] & $-$0.0093 & \phantom{$-$}0.143 \\
\hline
\end{tabular}
\mycaption{Leading fermionic three-loop corrections to EWPOs at $\mathcal{O}(\alpha^2\as)$ with \msbar\ prescription for the top mass.}
\label{tab:msbarewpos}
\end{table}
%-------------------------------------------------------------
\begin{table}[tb]
\centering
\renewcommand{\arraystretch}{1.2}
\begin{tabular}{|l | l | l | l | l|}\hline
 & \multicolumn{2}{c|}{on-shell $M_t$} & \multicolumn{2}{c|}{\msbar\ $m_\Pt$} \\
\hline
 & ${\cal O}(\alpha^2)$ & ${\cal O}(\alpha^2\as)$ & ${\cal O}(\alpha^2)$ & ${\cal O}(\alpha^2\as)$ \\
\hline
$\Delta r$ [$10^{-4}$] & \phantom{0}7.85 & $-1.09$ & \phantom{0}7.56 & $-0.50$\\
$\Delta\seff{f}$ [$10^{-5}$] & 30.98 & \phantom{$-$}1.31 & 31.18 & \phantom{$-$}0.75\\
$\Delta\overline\Gamma_\ell$ [MeV] & 0.2412 & $-0.0157$ & 0.2284 & $-0.0003$ \\
$\Delta\overline\Gamma_\nu$ [MeV] & 0.4145 & $-0.0002$ & 0.4152 & \phantom{$-$}0.0009 \\
$\Delta\overline\Gamma_{\rm d}$ [MeV] & 0.6666 & $-0.0049$ & 0.6780 & $-0.0018$ \\
$\Delta\overline\Gamma_{\rm u}$ [MeV] & 0.4964 & $-0.0203$ & 0.4911 & $-0.0029$ \\
$\Delta\overline\Gamma_{\rm tot}$ [MeV] & 4.951 & $-0.103$ & 4.947 & $-0.0093$ \\
\hline
\end{tabular}
\mycaption{Numerical comparison of leading fermionic $\mathcal{O}(\alpha^2)$ and $\mathcal{O}(\alpha^2\as)$ results between the on-shell and \msbar\ top-quark mass prescriptions. See text for more details.}
\label{tab:msbar}
\end{table}
%-------------------------------------------------------------

When using \msbar\ renormalization for the top-quark mass at ${\cal O}(\alpha^2\as)$, one must also use the \msbar\ top-quark mass, eq.~\eqref{mtmsbar}, as input at the lower order $\mathcal{O}(\alpha^2)$, for the sake of consistency. The leading fermionic $\mathcal{O}(\alpha^2)$ contributions have previously been computed in Refs.~\cite{mwshort,mwlong,swlept,swlept2,gz} and re-evaluated in Ref.~\cite{Chen:2020xzx}.
The corresponding numbers of both perturbative orders are listed in Table~\ref{tab:msbar}. One can see that the numerical changes at $\mathcal{O}(\alpha^2)$ and ${\cal O}(\alpha^2\as)$ partially compensate each other when going from the OS to the \msbar\ scheme. This is expected since the all-order results should be identical in both schemes (up to non-perturbative effects). The difference of the sum $\mathcal{O}(\alpha^2)+{\cal O}(\alpha^2\as)$ between the two schemes could be used as an estimate of the size of the unknown higher-order corrections at ${\cal O}(\alpha^2\as^2)$. A more detailed analysis of theoretical uncertainties from missing higher-order contributions will be left for future work.

%%%%%%%%%%%%%%%%%%%%%%%%%%%%%%%%%%%%%%%%%%%%%%%%%%%%%%%%%%%%%%

\section{Conclusions}
\label{conc}

This article reports on the calculation of mixed electroweak--QCD ${\cal O}(\alpha^2\as)$ corrections with two closed fermion loops to several important electroweak precision observables: the W-boson mass predicted from the Fermi constant, the partial and total decay widths of the Z-boson, and the effective weak mixing angle. On a technical level, this required the calculation of one- and two-loop self-energy integrals, as well as the derivation of the appropriate renormalization counterterms. To ensure gauge invariance, the complex pole scheme is adopted for the W- and Z-boson mass renormalization. Numerical results were presented for two definitions of the top-quark mass: in eqs.~\eqref{delros}--\eqref{dgamf} for the on-shell scheme, and in Tab.~\ref{tab:msbarewpos} for the \msbar\ scheme.

The numerical size of the corrections was found to be small compared to the experimental precision for direct measurements of these quantities today. However, they will be important for electroweak studies at future $e^+e^-$ colliders, such as CEPC, FCC-ee, or ILC/GigaZ. The order of magnitude of the corrections in the on-shell scheme matches expectations from previous estimates \cite{therr,Blondel:2018mad}. It is also observed that the magnitude of the leading fermionic ${\cal O}(\alpha^2\as)$ contributions is reduced when using the \msbar\ scheme for the top-quark mass. 

%Overall, the numerical ${\cal O}(\alpha^2\as)$ corrections are smaller than what one would estimate from naive counting of pre-factors, ${\cal O}(\alpha^2\as) \sim (N_cn_q+n_\ell)\alpha \times N_cn_q\alpha \times C_F\as \sim 4 \times 10^{-3}$, where $n_q$ and $n_\ell$ are the numbers of quark and lepton flavors, respectively, and $N_c=3$. This indicates that there are cancellations between  diagrams. Consequently, the remaining ${\cal O}(\alpha^2\as)$ corrections with only one closed fermion loop may be comparable in magnitude to the results presented in this article, and it will be important to compute them for the physics goals of future $e^+e^-$ colliders. These contributions require the evaluation of genuine 3-loop integrals and thus significant additional work.
%
On the other hand, one can observe that there are substantial numerical cancellations among the ${\cal O}(\alpha^2\as)$ corrections. For example, when setting $\Delta \alpha$ to zero the corrections to most EWPOs increase by roughly a factor of 5. This indicates that there are cancellations that are accidental in nature since they depend on the specific value of $\Delta \alpha$.
between terms involving $\Delta \alpha$ and other contributions. Consequently, the remaining ${\cal O}(\alpha^2\as)$ corrections with only one closed fermion loop may be comparable in magnitude to the results presented in this article, and it will be important to compute them for the physics goals of future $e^+e^-$ colliders. These contributions require the evaluation of genuine 3-loop integrals and thus significant additional work.

%%%%%%%%%%%%%%%%%%%%%%%%%%%%%%%%%%%%%%%%%%%%%%%%%%%%%%%%%%%%%%

\section*{Acknowledgments}

The authors would like to thank Yang Ma for useful discussions.
This work has been supported in part by the National Science Foundation under
grant no.\ PHY-1820760.

%%%%%%%%%%%%%%%%%%%%%%%%%%%%%%%%%%%%%%%%%%%%%%%%%%%%%%%%%%%%%%

\end{document}